# New methodology to design advanced MR-IR-UWB communication system

A. Lecointre, D. Dragomirescu and R. Plana

A new model is proposed giving the channel capability of a MB-IR-UWB system versus the number of subband and the duty cycle. The architecture simulated shows data rate ranging from 1,434 Gbits/s to 0.9 Gbits/s for 16 to 10 subbands and duty cycle ranging from 20% to 12%.

*Introduction*:

In the field of wireless sensor, one issue deals with the availability of wireless architectures featuring optimized performances in terms of power consumption and bit rate and lowering interference issues that has motivated the emergence of impulse radio ultra wideband medium (IR-UWB) [1] and more recently multiband IR-UWB (MB-IR-UWB) [2] aiming to have reconfigurable and flexible architecture that could be adapted following the application. However despite their attractive capabilities already demonstrated, there is a lack concerning a methodology aiming to design a MB-IR-UWB system under different requirements. This paper proposes an original approach to determine system design rules that will be useful to implement advanced communication architectures for wireless sensor network.

*Theoretical background*: In a low power and consequently low complexity context, IR-UWB techniques are suited as physical layer [1][2]. Assuming a receiver without equalization and binary modulation schemes to meet both simplicity of implementation, high data rate and low power consumption, the IR-UWB symbol duration is determined by the addition of the channel delay spread and the pulse duration [3]. From the Shannon relation defining the channel capacity *C* of an ideal band-limited channel *B* with additive white Gaussian noise (AWGN) interference [4]

$$C(\text{bits/s}) = B \cdot \log_2(1 + P_s/(B \cdot N_0)) = B \cdot \log_2(1 + \text{SNR}) \quad (1)$$

where $P_s$ is the signal power, SNR is the signal to noise ratio, and $N_0$ the noise spectral density, we can obtain the IR-UWB channel capacity. The expression of the latter includes parameters such as delay spread for dealing with time variant multipath channel. Since no inter-symbol interference (ISI) is assumed (to avoid equalization [3]) and binary modulation is imposed, the IR-UWB channel capacity can be expressed by only considering the temporal resolution defined by the channel delay spread $d_{spread}$, and the IR-UWB pulse duration $T_p$ :

$$C_{\text{IR-UWB}}(\text{bits/s}) = 1 / (T_p + d_{spread}) \quad (2).$$

Fig. 1 shows the IR-UWB channel capacity for binary modulation versus bandwidth in the case of channel delay spread of 10, 50, and 100 ns with a required SNR greater than 3 dB. Fig. 1 illustrates the existence of a delay spread asymptote at *1/$d_{spread}$* which limits the IR-UWB channel capacity. Near to the delay spread limitation, a bandwidth increase has a very low impact that motivates the use of multiband impulse ultra wide band technique.

*MB-IR-UWB channel capacity optimization*: Most of the time, MB-IR-UWB implementations, such as MB-OOK [5], use 500 MHz subbands and explain this bandwidth by the difficulty to generate an UWB pulse over 7.5 GHz, the reduction of data converter performance requirements, or the good behavior regarding flexibility for adaptation to local regulation. This widely used subband bandwidth of 500 MHz can also be explained by the UWB definition of the Federal Communications Commission (FCC), and by the implementation hardware constraints which limit the number of subbands and thus their size. However with the progress of the technology and continuous cost reduction, it could be possible to investigate different architectures and then opening some degree of freedom for designers that were not authorized before. Assuming implementation limitations, such as the maximum number of subbands $n_{max}$ (which is imposed by the number of mixers, the complexity and cost of the transceiver), the maximum bandwidth of a subband $Bs_{max}$ (which depends on the pulse generators and data converters performances), the total available bandwidth for the system $Bt_{max}$ (which is imposed by the local regulation) and the maximum authorized duty cycle $α_{max}$ of this discontinuous emission technique (IR-UWB), we express the bandwidth of each subband *Bs* and the number of subbands *n* by an optimization problem approach of the MB-IR-UWB capacity:

$$\max [C_{MB\text{-}IR\text{-}UWB}(Bs,n)] = n \times ( 1 / ( 1 / Bs + d_{spread} ) ) \quad (3)$$

with the following constraints :

$$n, n_{max} \text{ is a positive integer}$$

$$1 <= n < n_{max} \quad (4)$$

$$Bt = Bs \times n <= Bt_{max} \quad (5)$$

$$Bt / n_{max} \leq Bs \leq Bs_{max} \quad (6)$$

$$(1/α_{max} - 1) / d_{spread} \leq Bs. \quad (7)$$

Where *n* is the number of subbands which maximizes the channel capacity, $d_{spread}$ is the delay spread of the considered channel, *Bs* is the number of subbands for optimizing data rate and *Bt* is the entire bandwidth used by the IR-UWB system. The duty cycle that is a good indicator of the power consumption is defined by the following expression :

$$α = T_p / (T_p + d_{spread}) \quad (8)$$

Equations (3) to (7) indicate that implementation and channel constraints impact the achievable data rate of the MB-IR-UWB system. The optimization problem has a solution only if implementation constraints permit an achievable value for *Bs*. *Bs* must be located between $Bs_{min}$ and $Bs_{max}$. $Bs_{min}$ is defined by $n_{max}$, $α_{max}$, $d_{spread}$, and *Bt*:

$$Bs_{min} = \max [ Bt / n_{max} ; (1/α_{max} - 1) / d_{spread} ] \quad (9)$$

If $Bs_{max}$ is greater than $Bs_{min}$ then the problem has a solution. The value of *Bs*, which optimizes the MB-IR-UWB capacity under implementation constraints, is reached when the number of subband is the highest and when *Bs* is maximum for this optimized number of subbands.

Using the model and the methodology developed, we have performed simulations and the results are summarized in Figure 2. For these simulations, we have used a channel delay spread of 9 ns, a total available bandwidth $Bt_{max}$ of 7.5 GHz (FCC regulation), a $Bs_{max}$ of 750 MHz, $α_{max}$ equal to 20%, and $n_{max}$ equal to 30. $Bs_{max}$, $α_{max}$, $n_{max}$, are arbitrary chosen here to accommodate the hardware implementation constraints. This channel delay spread is extracted from the IEEE 802.15.4a UWB channel model in the case of industrial line of sight environment [6]. First of all, we can observe that the MB-IR-UWB capacity ranges from 1.434 Gbits/s to 0.967 Gbits/s when the subband bandwith varies from 444,4 MHz to 750 MHz. It has to be outlines that the maximum data rate (1434 Mbits/s) is achieved for the largest number of subband (16) and the locally largest subband bandwidth (464 Mhz). This maximum corresponds also to the locally lowest power consumption (since the duty cycle is at its locally lowest value 19.3 %). Our model also provides the subband channel capacity that varies from 88 Mbits/s to 97Mbits/s when the subband ranges from 444,4MHz to 750 MHz. Finally, the simulations give indication on the duty cycle that is obtained versus the subband bandwidth ranging from 20% to 12% versus subband bandwidth.

*Perspectives and conclusions:* This paper proposes an original approach to calculate the capabilities of MB-IR-UWB channel and to exploit this approach to design the future communication architecture that will be implemented within wireless sensor network where it is often necessary to accommodate both high data rate and low power consumption. The model proposes new design rules involving channel capacity, duty cycle versus number of subband and subband bandwidth that allow designing a MB-IR-UWB architecture featuring data rate ranging from 1.434 Gb/s to 1 Gb/s with number of subband ranging from 16 to 10 and duty cycle varying from 20% to 12% over a 9ns delay spread UWB channel. This open the way of a new generation of software radio with enhanced capabilities.

**Authors' affiliations:**
A. Lecointre, D. Dragomirescu and R. Plana (University of Toulouse, LAAS-CNRS, 7 Avenue du Colonel Roche, 31077 Toulouse Cedex 4, FRANCE)
alecoint@laas.fr, daniela@laas.fr, plana@laas.fr.

**Figure captions:**

Fig. 1  IR-UWB channel capacity versus bandwidth (required SNR > 3 dB)
——□——     delay spread = 50 ns
    □         1 / 50 ns
——o——     delay spread = 20 ns
    o         1 / 20 ns
——•——     delay spread = 10 ns
    •         1 / 10 ns

Fig. 2 Simulations of the MB-IR-UWB channel capacity (2.b), duty cycle (2.b), number of subband (2.a) and subband channel capacity (2.a) versus subband bandwith using the model developed.

Figure 1

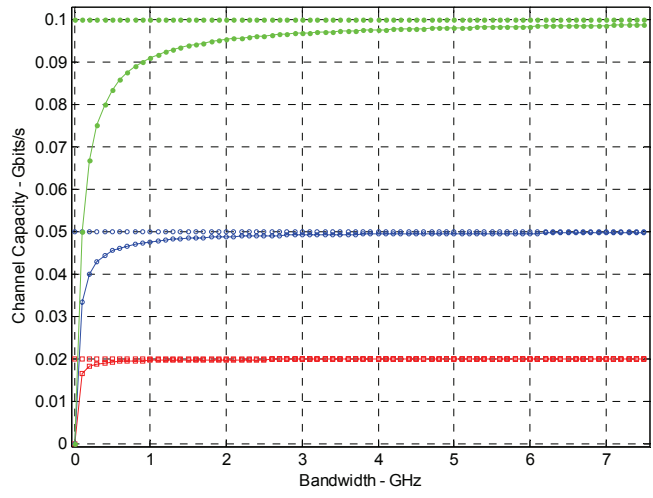

Figure 2a

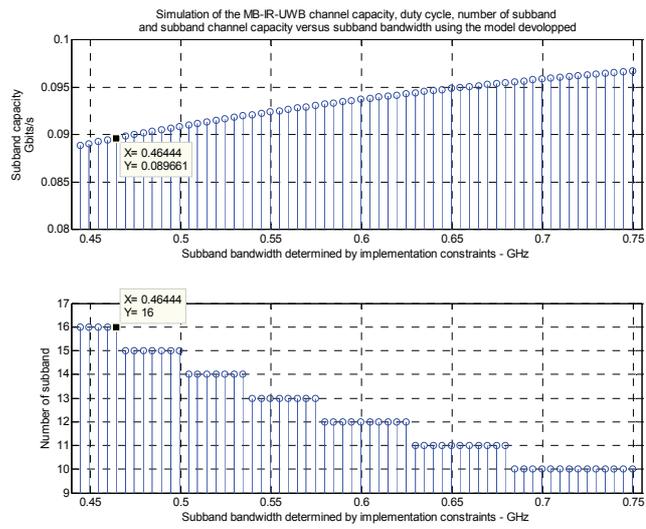

# Figure 2b

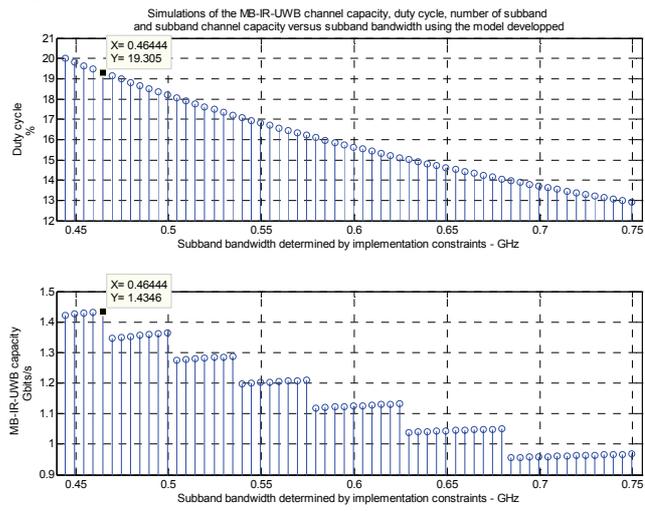